\documentclass[11pt,a4paper]{article}

\usepackage{geometry}
\usepackage[applemac]{ inputenc}%Usar este pacote de acento no mac

\usepackage{mathrsfs}
\usepackage{amsmath}
\usepackage{amssymb}
\usepackage{multicol}
\usepackage{float}
\usepackage{makeidx}
\usepackage{layout}
\usepackage{array}
\usepackage{hyperref}
\usepackage{latexsym}
\usepackage{indent first}

\usepackage{subcaption,graphicx}
\usepackage{appendix}

\usepackage{color}

\usepackage[sectionbib,square]{natbib}%Pacote para definir novas formas de citacao

\usepackage[normalem]{ulem}%Pacote para sublinhar o texto de diversos modos

\DeclareMathOperator{\sech}{sech}
\title{Gravity-capillary flows over obstacles for the fifth-order  forced  Korteweg-de Vries equation}
\author{M. V. Flamarion$^{1}$ and R. Ribeiro-Jr$^{2}$}
\date{}
\begin{document}

\maketitle

{\footnotesize
	\begin{center}
	$^{1}$ UFRPE/Rural Federal University of Pernambuco, UACSA/Unidade Acad{\^e}mica do Cabo de Santo Agostinho, BR 101 Sul, 5225, 54503-900, Ponte dos Carvalhos, Cabo de Santo Agostinho, Pernambuco, Brazil.
	
		$^2$ UFPR/Federal University of Paran\'a,  Departamento de Matem\'atica, Centro Polit\'ecnico, Jardim das Am\'ericas, Caixa Postal 19081, Curitiba, PR, 81531-980, Brazil. 
	
	\end{center}
	
	}
\begin{abstract}
The aim of this work is to investigate gravity-capillary waves resonantly excited by two topographic obstacles in a shallow water channel.  By considering  the weakly nonlinear regime the forced  fifth-order Korteweg-de Vries equation arises as a model for the free surface displacement. The water surface is initially taken at rest and the initial value problem for this equation is computed numerically using a pseudospectral method. We study near-resonant flows with intermediate capillary effects. Details of the wave interactions are analysed for obstacles with different sizes. Our numerical results  indicate that the flow is not necessarily governed by the larger obstacle.   \\

 
\end{abstract}

\section{Introduction}

Waves excited by an external force is of great
current interest due to the large number of  physical applications.  For instance, nonlinear eletrical lines,  superconductive electronics, elementary-particle physics (\cite{ Peyrard, Joseph}) and hydrodynamics. Regarding the last one,  we mention, flow of water over rocks, ship waves \cite{Baines}, and waves generated by storms \cite{Johnson}. In water waves, the external force usually models a pressure distribution or a topographic obstacle.

In the absence of surface tension,  the fundamental parameter used for describing the pattern of  generated waves due to a current-topography interaction is the Froude number $$F = \frac{U_{0}}{\sqrt{gh_{0}}}.$$ 
Here, $U_0$ is the velocity of the uniform stream, $g$ is the acceleration of gravity and $h_{0}$ is the undisturbed depth of a shallow water channel. The Froude number is critical when $F= 1$, i.e., 
when the linear long-wave phase speed is equal to the mean flow speed. In the weakly nonlinear regime the forced Korteweg-de Vries (fKdV) model is valid to study near-resonant flows ($F\approx 1$) over obstacles with small amplitudes. 
 A detailed study considering the fKdV equation was first done by \cite{Wu2},  later by \cite{Akylas, Grimshaw86, Wu1, Paul, Ermakov}, and more recently by \cite{Marcelo-Paul-Andre} for a vertically sheared current. 
 All these authors have considered only one obstacle.  
 
Regarding  flow over multiple obstacles, \cite{Chardard} investigated numerically the stability of solitary waves.  \cite{Lee1}  studied trapped waves between two obstacles. 
 They considered a bottom topography with two bumps and found numerical solutions for the fKdV which remained bouncing back and forth between the  obstacles for a certain period of time. %These authors  computed numerical steady waves for the fKdV equation and used their perturbations as initial datas. \
 \cite{Grimshaw16} considered a near-resonant flow over two obstacles. They defined the development of the flow in three stages. 
The first stage is characterized by the formation of an undular bore over each obstacle. The second is the interaction of the generated waves between the obstacles, and the third is  the evolution at large times when the larger  obstacle controls the flow. More recently, these authors studied the interaction of generated waves over two obstacles (bumps and holes) in the near-resonant
regime describing the dynamic of the wave interactions in great details  (\cite{Grimshaw19}). 

When the surface tension is included  in the problem, an additional parameter becomes fundamental in the study of generated waves, namely the Bond number $$B = \frac{\sigma}{\rho g h_{0}^2},$$ where $\sigma$ is the coefficient of the surface tension and $\rho$ is  the constant density of the fluid.  Gravity-capillary waves can also be described by the Korteweg-de Vries (KdV) equation. However, the dispersion term in the quation vanishes when $B$ is critical, i.e, $B=1/3$. Thus,  solitary wave  solutions ($\sech^2$-like) are no longer appropriate since the length scale of these waves becomes zero (\cite{Falcon}). Studying flows over obstacles under gravity-capillary effects, \cite{Milewski3} derived a fifth-order fKdV for $F\approx 1$ and $B\approx 1/3$. They showed that this equation has unsteady solitary waves solutions with small oscillating tails.  A numerical investigation of solitary waves and collisions for the fifth-order KdV was done by \cite{Malomed}.  They found solitary waves with oscillatory tails, and when two of these waves interact  some of them regain their shape while others split into several solitary waves of different types.

Recently, \cite{Hirata} used the body-fitted curvilinear coordinates to solve Euler's equations in the presence of an obstacle with a uniform flow and compared the results with the fKdV and the  fifth-order-fKdV for the resonant flow ($F=1$) and intermediate capillary effects ($B\approx 1/3$).  They observed short wave radiation  when the effects of surface tension are lesser $(B <1/3)$. Besides, a train of solitary waves propagating upstream radiates short linear waves whose phase speed is equal to the upstream-advancing speeds of the solitary wave. 
The fifth-order-fKdV captured the wave train propagating upstream, however it predicted waves of longer length, which is natural since the KdV-models are based on a long wave approach.

In this paper we  investigate numerically  the interaction of excited gravity-capillary waves in the near-resonant flow over two obstacles for the fifth-order-fKdV. More precisely, we focus on the case in which the Froude number is near critical  and the capillary effects are intermediate. The problem is studied for obstacles with different sizes. To the best of our knowledge there are no articles regarding the fifth-order-fKdV in the presence of two obstacles.   From our experiments, we identify that there are regimes in which the flow is not necessarily driven by the larger obstacle, what is different from the case in which surface tension is neglected (\cite{Grimshaw16, Grimshaw19}). Besides, we present in details the main features of the near-resonant flow. 

This article is organized as follows. In section 2 we present the mathematical formulation of the non-dimentional fifth-order-fKdV equation.  The numerical results are presented in section 3 and the conclusion in section 4.

\section{The fifth-order forced Korteweg-de Vries equation}
We consider a two-dimensional incompressible and irrotational flow of an inviscid fluid with constant density ($\rho$) in a shallow water channel of undisturbed depth ($h_0$) and in the presence of an uniform flow ($U_0$). Besides,  the fluid is under the gravity force ($g$) and the surface tension ($\sigma$). 

In the weakly nonlinear regime, the dimensionless forced fifth-order Korteweg-de Vries (5th-order-fKdV)
\begin{equation}\label{fKdV}
\zeta_{t}+f\zeta_{x}-\frac{3}{2}\zeta\zeta_{x}+\frac{b}{2}\zeta_{xxx}-\frac{1}{90}\zeta_{xxxxx}=\frac{1}{2} h_{x}(x),
\end{equation}
is used to describe the flow over small obstacles (\cite{Zhu, Milewski3, Hirata}). Here, $\zeta(x,t)$ is the free-surface displacement over the undisturbed surface and $h(x)$ is the obstacle submerged. 
The parameter $f$ represents a perturbation of the Froude number, i.e, $F=1+\epsilon f$, and $b$ is a perturbation of the Bond number  $B=1/3+\epsilon^{1/2}b$, where $\epsilon>0$ is a small parameter. The flow is  supercritical, subcritical or near-resonant depending on whether $f > 0$, $f < $ or $f \approx 0$.  Analogously, the capillary effect is strong, weak or intermediate whether $b > 0$, $b<0 $ or $b \approx 0$.

The 5th-order-fKdV equation (\ref{fKdV}) is solved numerically using a Fourier pseudospectral method with an integrating factor. It avoids numerical instabilities due to the higher-order dispersive term. We consider the computational domain  with a uniform grid. All derivatives in $x$ are performed spectrally (\cite{Trefethen}).  In addition, the time evolution is computed through the Runge-Kutta fourth-order method (RK4). The initial wave profile is always taken at rest $(\zeta(x,0)=0)$. Since the fKdV fails when the Bond number is critical, we focus exactly on this case ($b=0$).

\section{Numerical results}

\begin{figure}[!ht]
	\centerline{\includegraphics{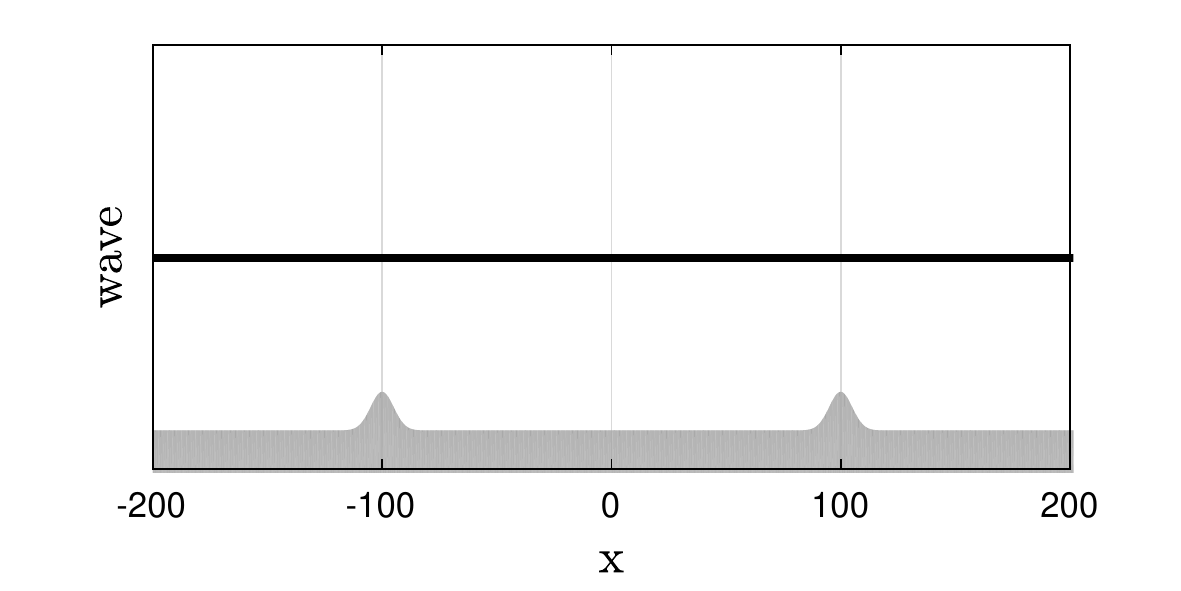}}
	\caption{Sketch of the physical problem at $t=0$.}
	\label{fig:S.0}
\end{figure}

In the same fashion as presented in \cite{Grimshaw19}, we consider a bottom topography modelled by two localised obstacles
\begin{equation}\label{obstacles}
h(x) = \epsilon_{1}\exp\Big({-(x-x_a)^2/w}\Big)+\epsilon_{2} \exp\Big({-(x-x_b)^2/w}\Big),
\end{equation}
%\begin{equation}\label{obstacles}
%h(x) = \epsilon_{1}\sech \big(k(x-x_a)\big)^2+\epsilon_{2}\sech \big(k(x-x_b)\big)^2,
%\end{equation}
where $\epsilon_1$ and $\epsilon_2$ are the amplitudes of the obstacles,  $w$ is the width of the obstacles, and $x_a$ and $x_b$ are their locations. We focus on the cases in which $\epsilon_{1}$ and $\epsilon_{2}$ are positive and let the parameter $f$ vary. There are a long list of parameters to be considered so that we fix $x_{a}=-100$, $x_{b}=100$, and $w=50$. This choice of parameters let us observe waves being generated over the obstacles independently at early times, and later  analyse their interactions.  A sketch of the physical problem at $t=0$ is depicted in Figure \ref{fig:S.0}. Since the current is turned on at $t=0^{+}$ waves are immediate generated.

In the following subsection we present our results in the non-resonant and near-resonant regimes.
%\subsection{$\epsilon_1=0.01$ and $\epsilon_2=0.01$}

\subsection{Non-resonant regime}
In this regime we let $|f|$ be sufficiently large and investigate how changes of the amplitude of the obstacles affect the flow. 

\subsubsection*{Supercritical regime}
In a presence of one single obstacle, the supercritical non-resonant flow is characterized by the formation of an elevation wave over the obstacle and depression solitary waves propagating downstream. Besides, when $t\rightarrow\infty$,  the steady state is reached after a train of downstream waves move away from the obstacle. Radiation of short waves is not observed (\cite{Milewski3}).

\begin{figure}[!ht]
	\centerline{\includegraphics{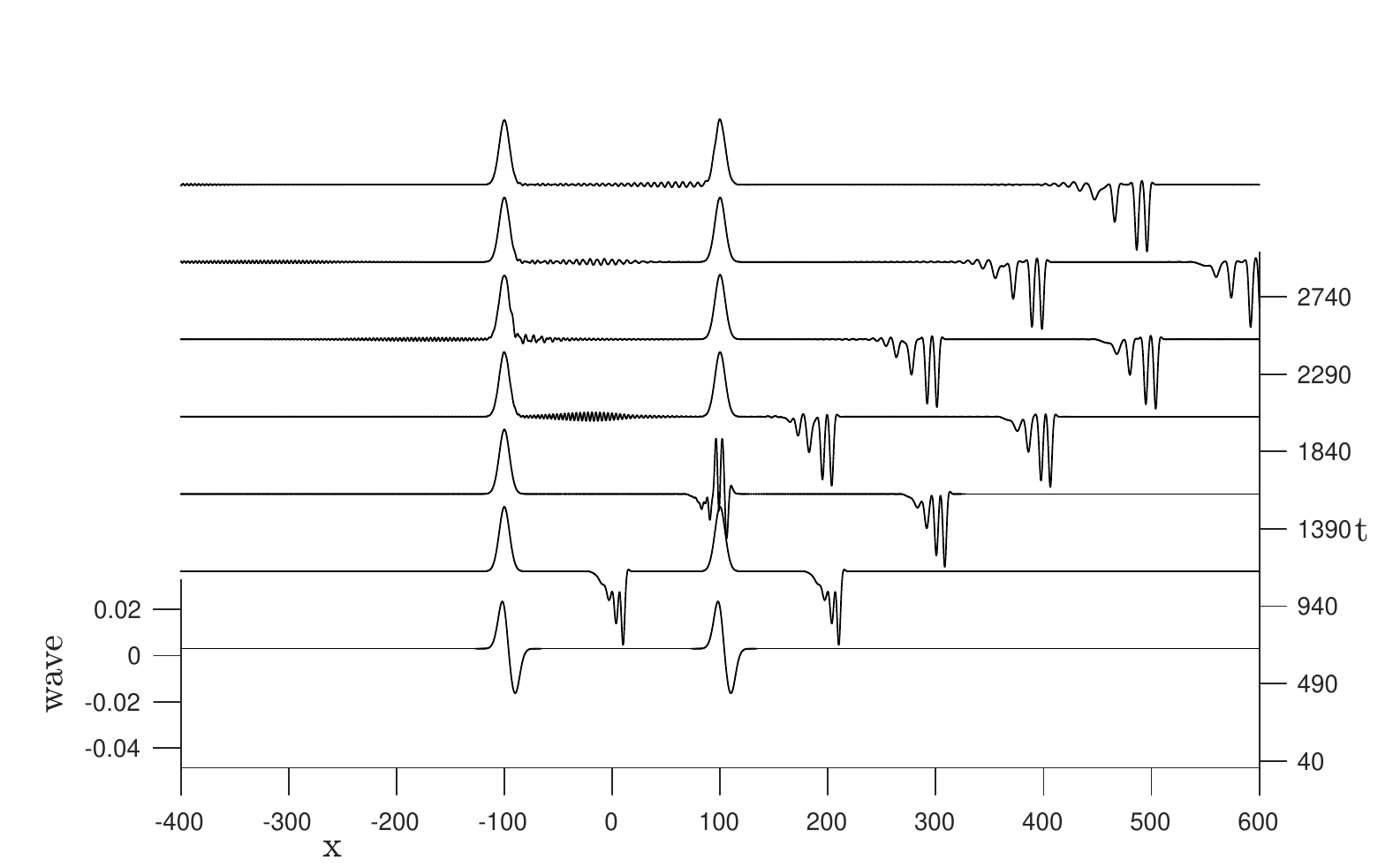}}
	\caption{Supercritical non-resonant regime free-surface evolution with  $f=0.2$, $\epsilon_{1}=0.01$ and $\epsilon_{2}=0.01$.}
	\label{fig:S.1}
\end{figure}

We first consider two obstacles with same amplitudes $\epsilon_{1}=\epsilon_{2}=0.01$. In this case, we observe depression solitary waves propagating downstream and the formation of two elevation waves over the obstacles. These elevation waves no longer reach the steady state because radiation of short waves is emitted from the right obstacle to the left one. Part of the radiation is reflected back remaining between the bumps and the other moves away from both obstacles. Since capillary effects are under consideration, short waves travel faster,  therefore, the downstream solitary waves are constantly disturbed by the radiation.  Figure \ref{fig:S.1} illustrates this flow motion.

\begin{figure}[!ht]
	\centerline{\includegraphics{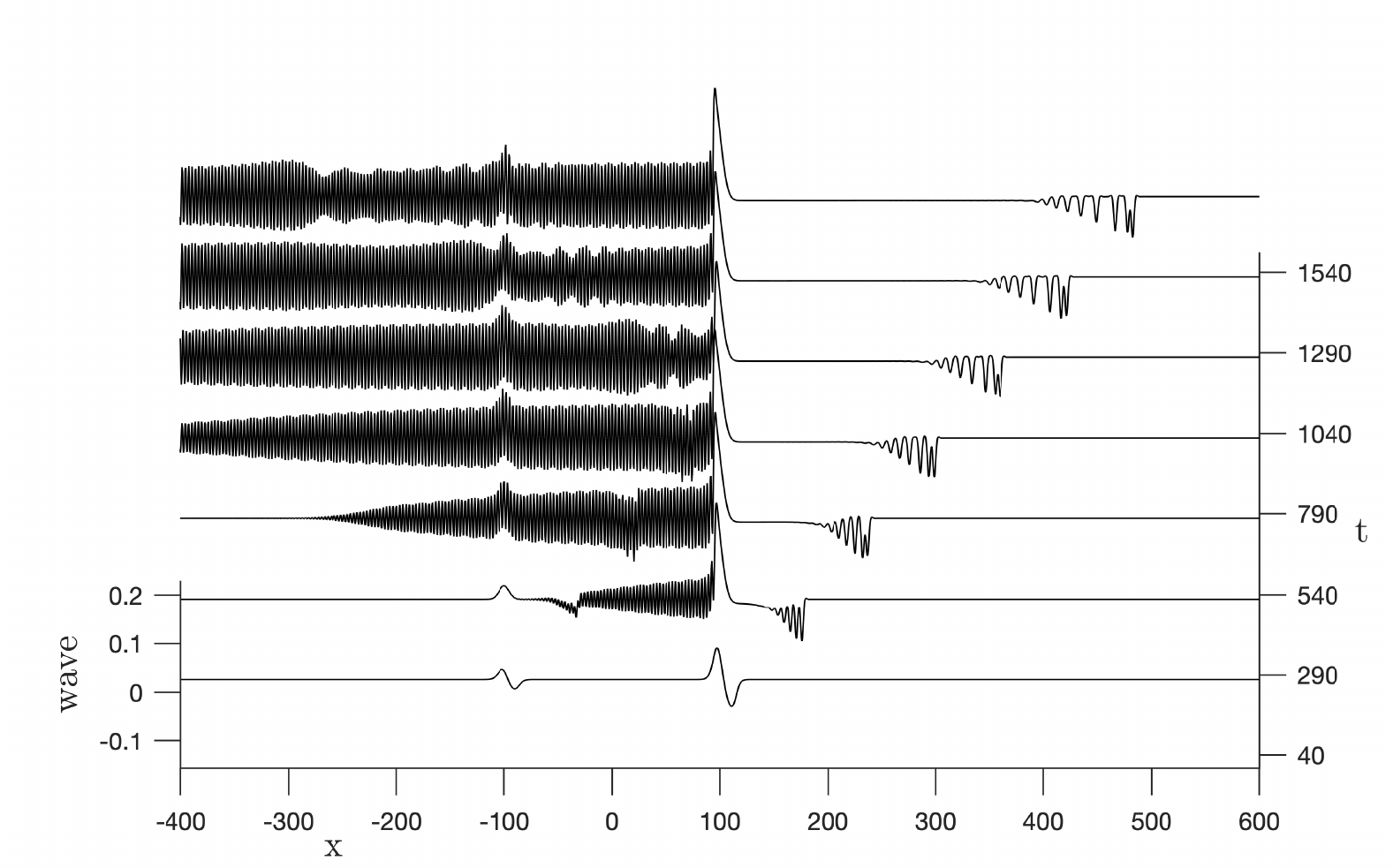}}
	\caption{Supercritical non-resonant regime free-surface evolution with  $f=0.2$, $\epsilon_{1}=0.01$ and $\epsilon_{2}=0.03$.}
	\label{fig:S.2}
\end{figure}

Now, we increase the amplitude of the second obstacle by considering $\epsilon_{2}=0.03$. In this regime, the flow is governed by the larger obstacle as can be seen in  Figure \ref{fig:S.2}. At early times, an elevation wave forms above both  obstacles, as time goes on a wave train emitted from the larger obstacle swallows the wave over the smaller one destroying the expected steady elevation wave. Differently from the previous case,  depression solitary waves propagate downstream without radiation being observed. Waves are trapped between the obstacles with part of it being radiated upstream.

\begin{figure}[!ht]
	\centerline{\includegraphics{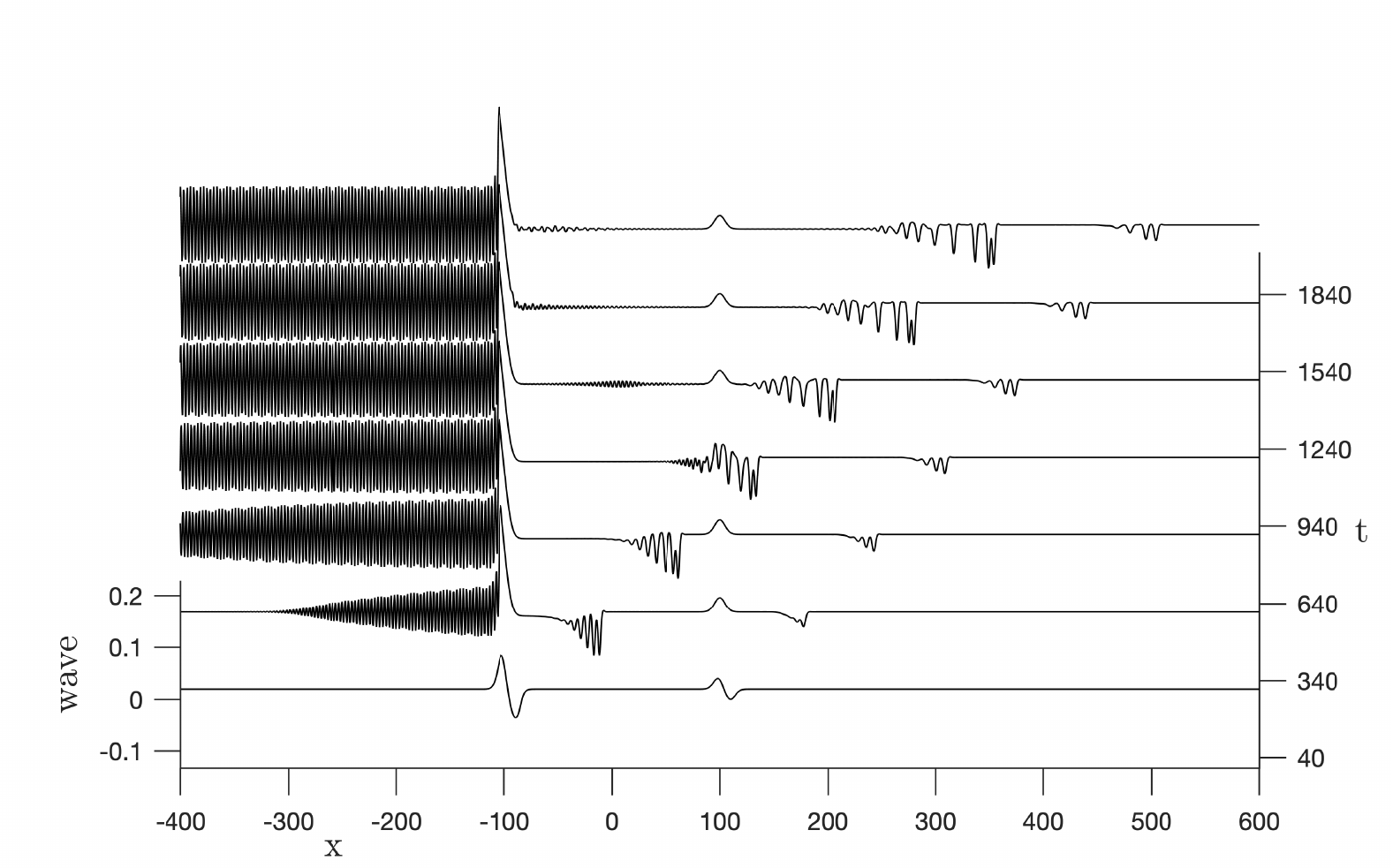}}
	\caption{Supercritical non-resonant regime free-surface evolution with  $f=0.2$, $\epsilon_{1}=0.03$ and $\epsilon_{2}=0.01$.}
	\label{fig:S.3}
\end{figure}

Lastly,  we choose $\epsilon_{1}=0.03$ and  $\epsilon_{2}=0.01$. The dynamic is just a simple reflection from the previous case at early times. A elevation wave is generated over the two obstacles with depression solitary waves being emanated from both obstacles, being larger the ones that come from the larger  obstacle. These waves pass the smaller obstacle which radiates short waves upstream. Once the radiation reaches the larger obstacle, it reflects part of it and the elevation wave is no longer steady. This situation is displayed in details in Figure \ref{fig:S.3}. We see that radiation moves back and forth between the bumps with some portion moving away from the obstacles. Although the elevation wave over the smaller obstacle is not steady, it is not destroyed as in the previous case. 
\subsubsection*{Subcritical regime}
In the one-obstacle problem, the subcritical non-resonant flow is mainly described for the formation of a  depression solitary wave over the obstacle and a wave train propagating upstream. Moreover,  when $t\rightarrow\infty$, the solution of (\ref{fKdV}) reaches a steady state, which in this regime is a free-surface depression wave above the obstacle (\cite{Milewski3}).

\begin{figure}[!ht]
	\centerline{\includegraphics{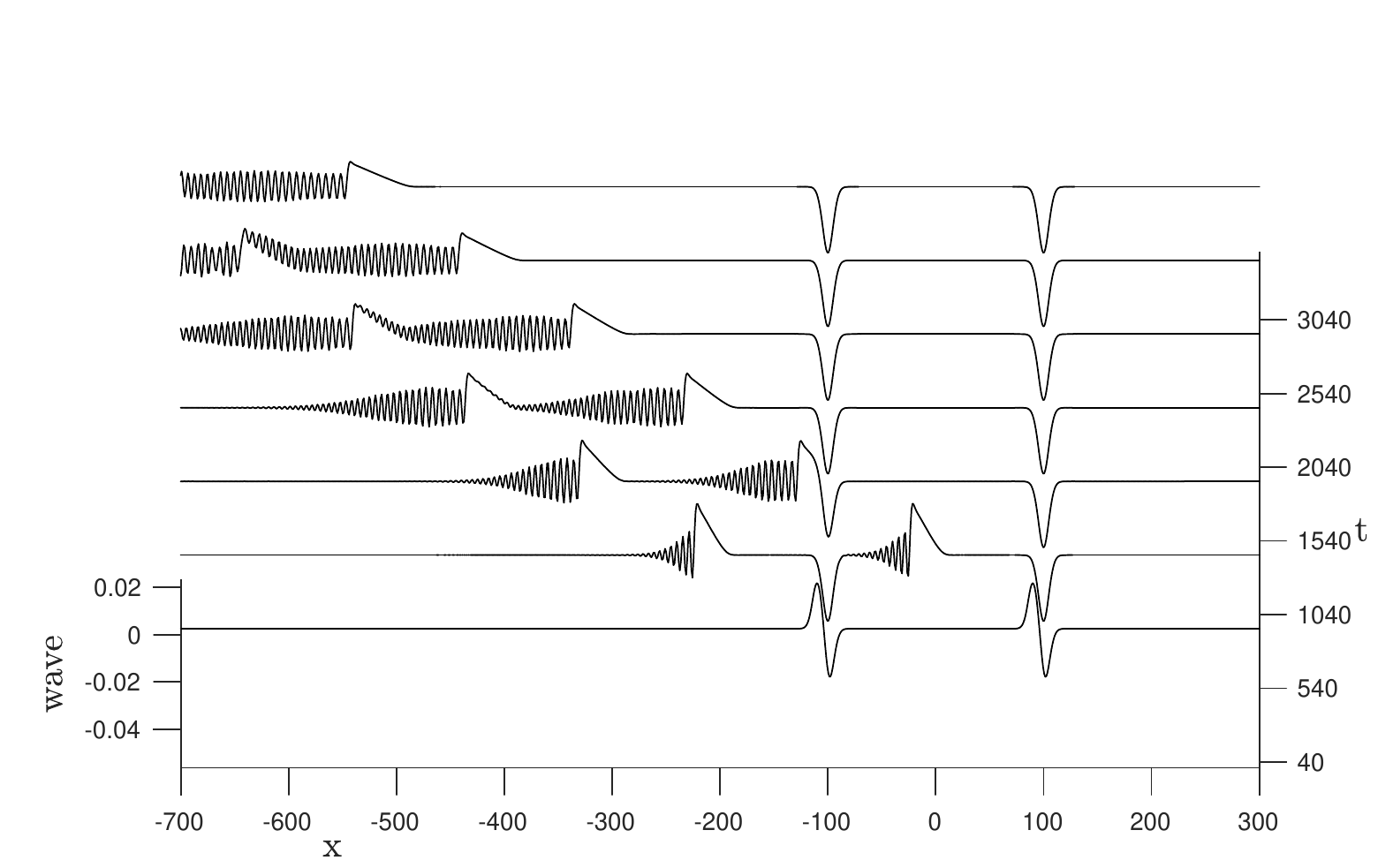}}
	\caption{Subcritical non-resonant regime free-surface evolution with  $f=-0.2$, $\epsilon_{1}=0.01$ and $\epsilon_{2}=0.01$.}
	\label{fig:S.4}
\end{figure}

In order to study the two-obstacle problem, we initially fix $\epsilon_{1}=\epsilon_{2}=0.01$. In this scenario the dynamic of the free surface behaves as the following: at early times, an elevation wave is generated above both obstacles and wave trains are emanated upstream. Then the wave train generated  from the right obstacle passes over the left one, gains kinect energy and interacts with the other wave train. The depression waves over the obstacles remain intact. The steady state is obtained when the upstream train waves are shed. Figure \ref{fig:S.4} illustrates this dynamic. We observe that the short wave train generated from the right obstacles overtakes the one from the left one.

\begin{figure}[!ht]
	\centerline{\includegraphics{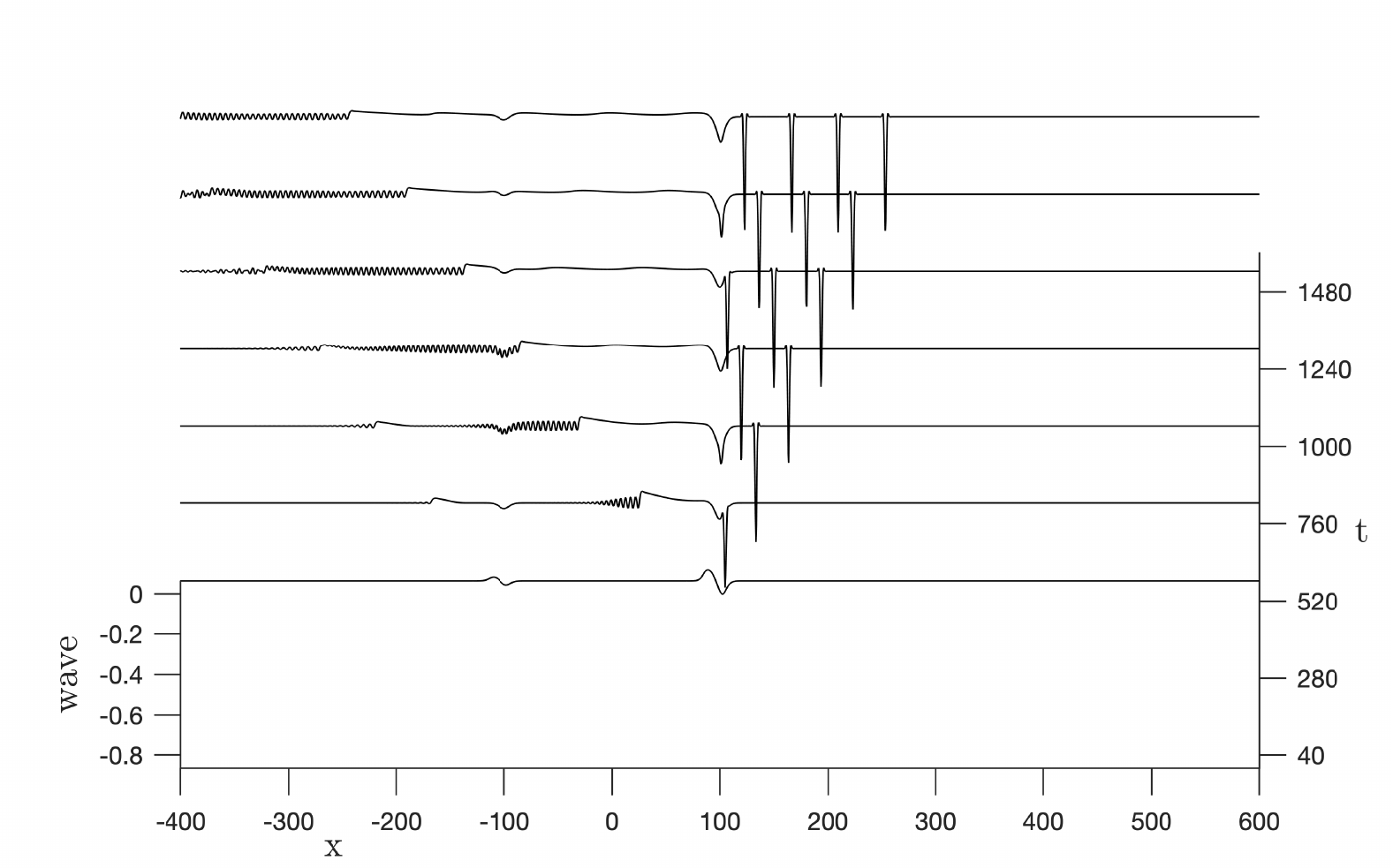}}
	\caption{Subcritical non-resonant regime free-surface evolution with  $f=-0.2$, $\epsilon_{1}=0.01$ and $\epsilon_{2}=0.03$. }
	\label{fig:S.5}
\end{figure}

For $\epsilon_{1}=0.01$ and $\epsilon_{2}=0.03$ the flow is quite different as depicted in Figure \ref{fig:S.5}. Depression solitary waves are now emitted periodically downstream from the larger obstacle. Steady waves no longer occur. An undular bore forms between the bumps with a wave train ahead. This wave train interacts with the elevation wave formed above the smaller obstacle destroying its shape. As time elapses, an elevation wave rises again, but it is no longer a steady wave because an undular bore remain between the obstacles. 

\begin{figure}[!ht]
	\centerline{\includegraphics{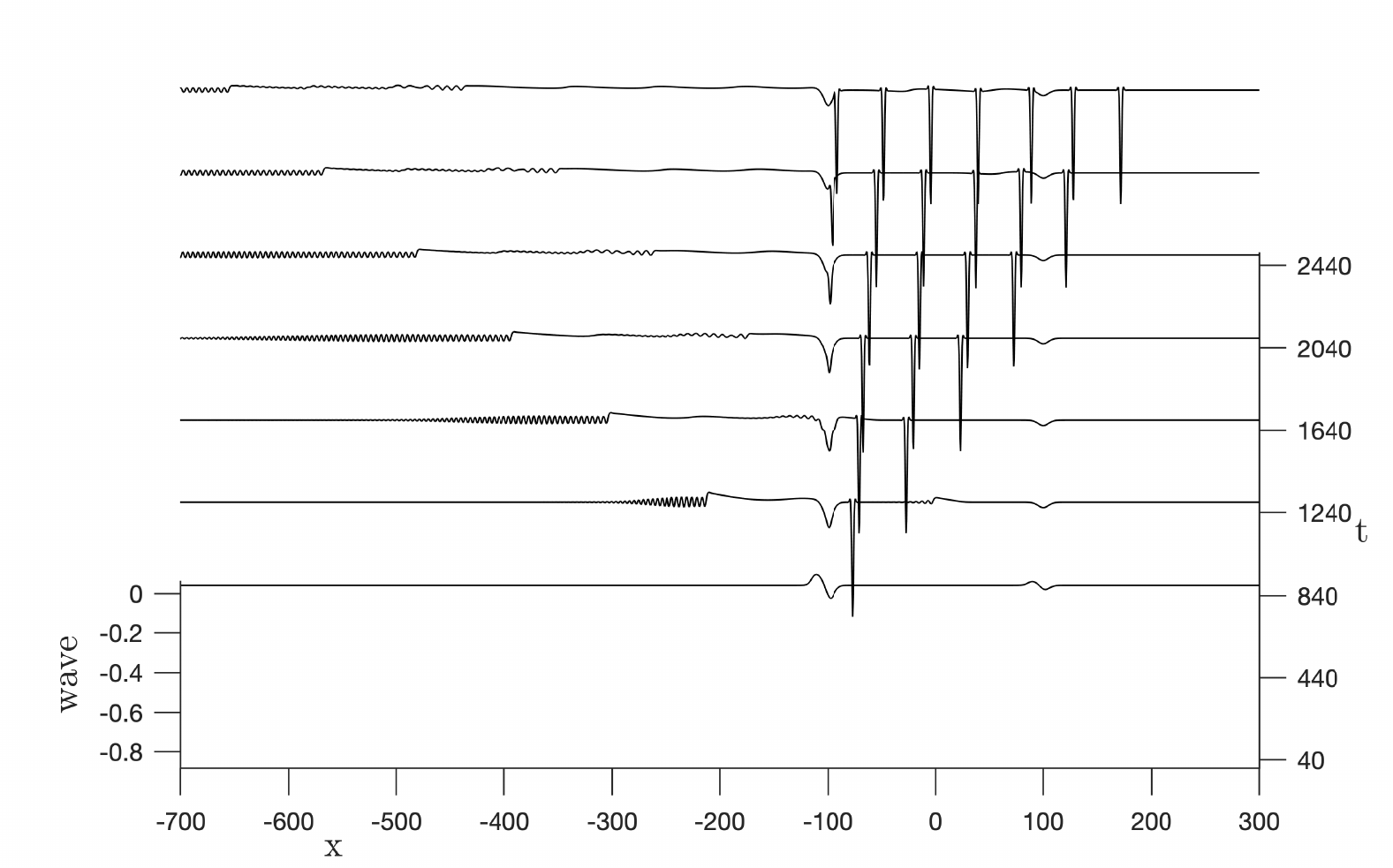}}
	\caption{Subcritical non-resonant regime free-surface evolution with  $f=-0.2$, $\epsilon_{1}=0.03$ and $\epsilon_{2}=0.01$.}
	\label{fig:S.6}
\end{figure}

Now, we reverse the role of the obstacles by taking $\epsilon_{1}=0.03$ and $\epsilon_{2}=0.01$. As in the previous case the larger obstacle generates depression solitary waves periodically which propagates downstream. This generation radiates small short waves upstream.  The depression solitary waves have enough energy to overcome the small obstacle. Although their shape is changed during the passage over the second obstacle, after the interaction it regains its form. This dynamic is displayed in Figure \ref{fig:S.6}. The initial steady elevation waves above the obstacles no longer persist. 

\subsection{Near-resonant regime}
In this regime we let $f$ be closer to $0$ and consider the same types of obstacles as before. As we will  show the wave interaction is somehow more nonlinear, so differently from the non-resonant regime it is difficult to capture a pattern in the generated waves. %In supercritical regimes we found that the flow is mainly characterised by an undular bore formation between the obstacles starting from the larger obstacle. In the subcritical regimes the main feature observed is the generation of depression solitary waves dowsntream. There are regimes in which waves are trapped between the bumps.
\begin{figure}[!ht]
	\centerline{\includegraphics{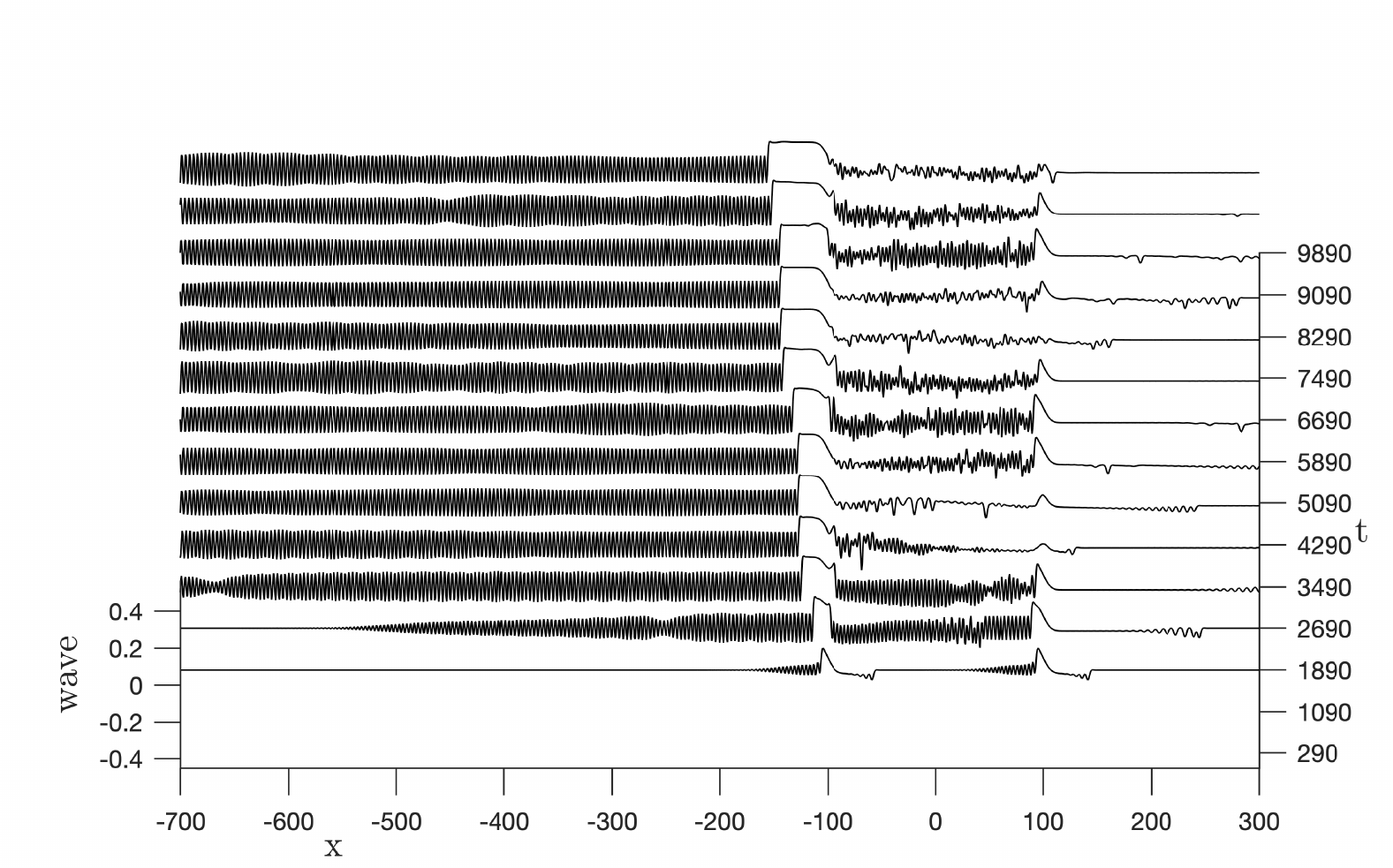}}
	\caption{Supercritical near-resonant regime free-surface evolution with  $f=0.1$, $\epsilon_{1}=0.01$ and $\epsilon_{2}=0.01$.}
	\label{fig:S.7}
\end{figure}
\subsubsection*{Supercritical regime}
We start  choosing $\epsilon_{1}=0.01$ and $\epsilon_{2}=0.01$. In this case, we observe a formation of an undular bore above the left obstacle. It propagates very slowly  upstream led by a  series of wave trains. In the region between the obstacles, at early times, wave trains propagate upstream. Part of the wave trains are reflected from the first obstacle and part of it radiates. So, radiation is observed between the bumps for large times. Above the right obstacle, we notice an unstable elevation wave.  Figure \ref{fig:S.7} pictures this scenario. Although the dynamic between the obstacles is very unpredictable the formation of the undular bore is very clear.

\begin{figure}[!ht]
	\centerline{\includegraphics{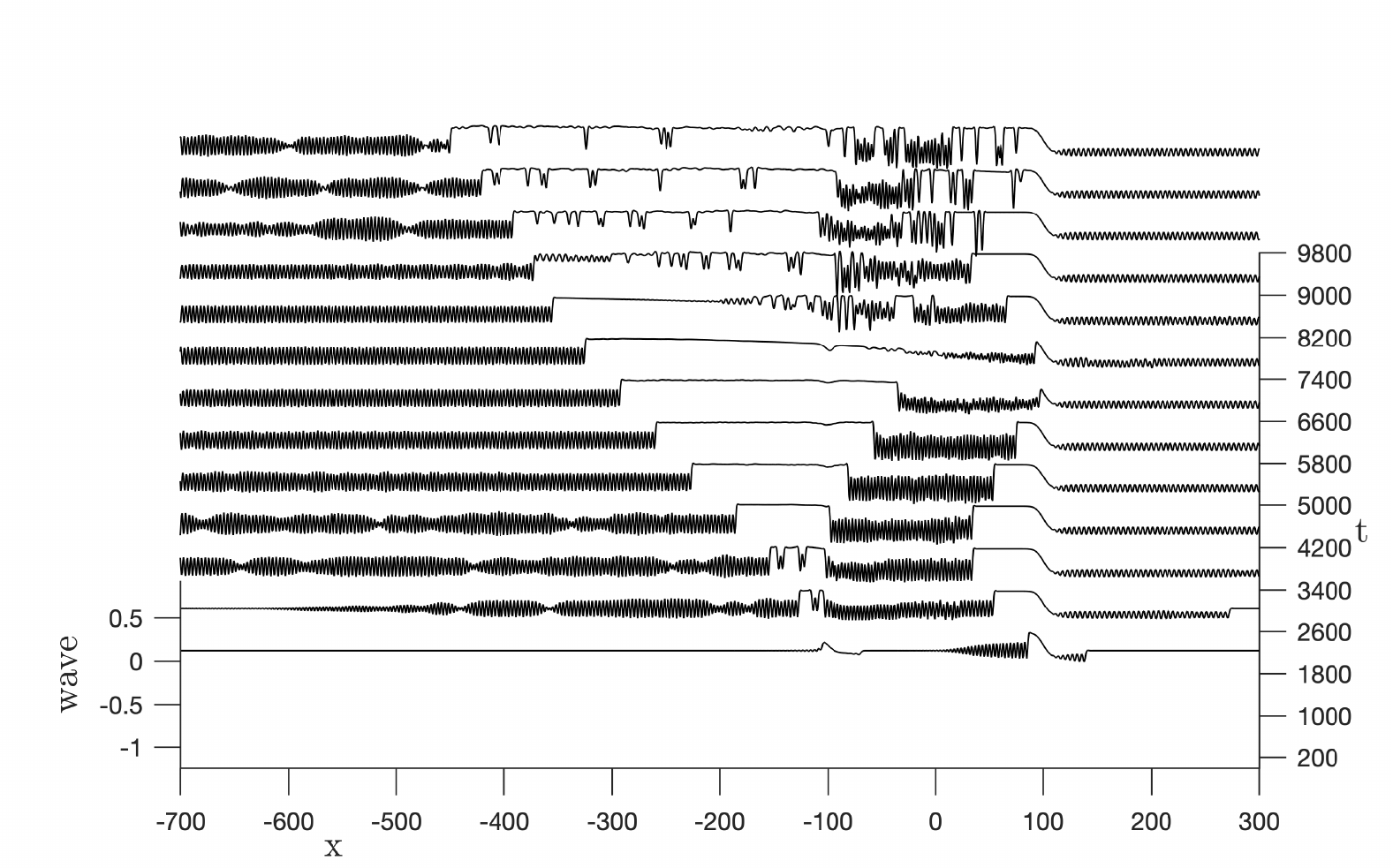}}
	\caption{Supercritical near-resonant regime free-surface evolution with  $f=0.1$, $\epsilon_{1}=0.01$ and $\epsilon_{2}=0.03$. }
	\label{fig:S.8}
\end{figure}

When the second obstacle is larger, $\epsilon_{1}=0.01$ and $\epsilon_{2}=0.03$, the flow structure is very distinct. A formation of an undular bore  is observed above both obstacles. The one over the small obstacle propagates upstream and downstream while the other upstream. As a consequence of that, wave trains accumulate between the bumps. Later, the accumulated wave trains is compressed and then the undular bore colapses. After a while,  waves start moving upon the bore. Figure \ref{fig:S.8} depicts the situation described. Around $t=5800$ the undular bore starts deteriorating and waves run over it. It is important to notice that it is not clear which obstacle controls the flow for large times. If we look at Figure \ref{fig:S.8} for $t>8200$ we cannot point which obstacle is larger, what is different from  the supercritical non-resonant regime (see Figure \ref{fig:S.2}). %it is clear that the larger obstacle controls the criticality of flow (see Figure \ref{fig:S.2}).

\begin{figure}[!ht]
	\centerline{\includegraphics{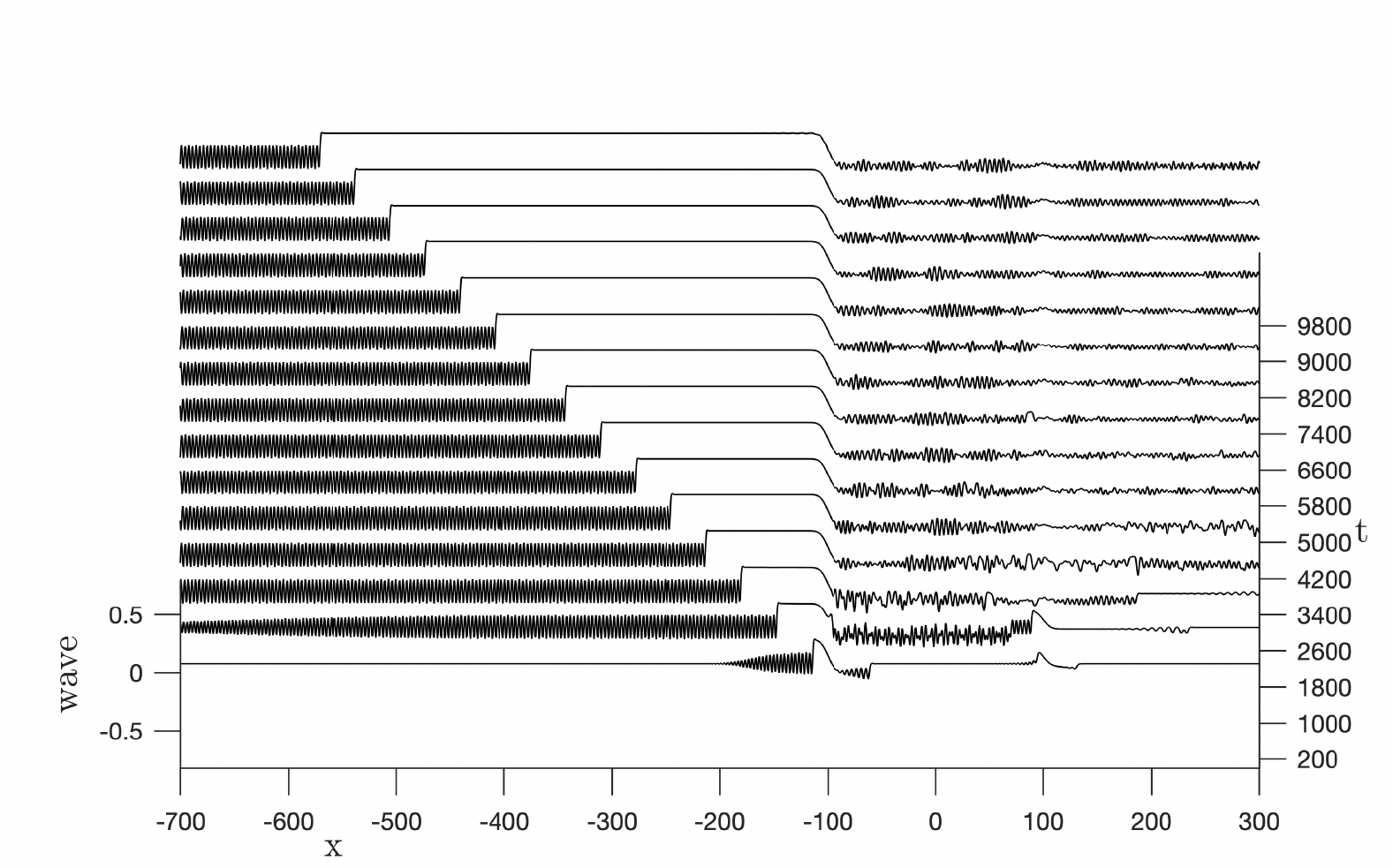}}
	\caption{Supercritical near-resonant regime free-surface evolution with  $f=0.1$, $\epsilon_{1}=0.03$ and $\epsilon_{2}=0.01$.}
	\label{fig:S.9}
\end{figure}

Now, let us consider the case where the first obstacle is larger, namely $\epsilon_{1}=0.03$ and $\epsilon_{2}=0.01$. This case is much clear than the previous one. We see the formation of an undular bore above the larger obstacle and radiation between the obstacles which is expelled downstream. Figure \ref{fig:S.9} shows the free-surface evolution for $f=0.1$. The flow is driven by the larger obstacle.

\subsubsection*{Subcritical regime}
\begin{figure}[!ht]
	\centerline{\includegraphics{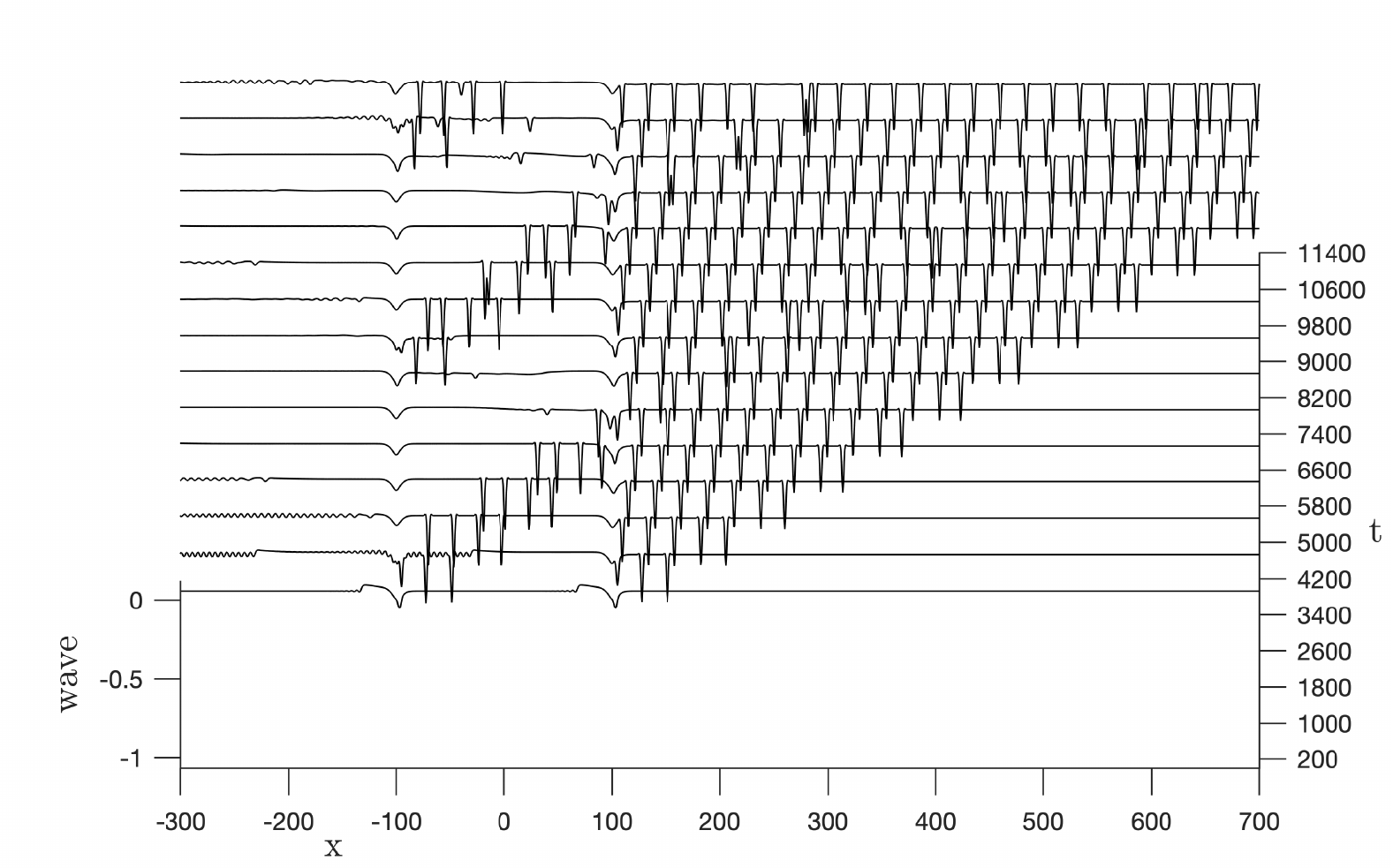}}
	\caption{Subcritical resonant regime free-surface evolution with  $f=-0.1$, $\epsilon_{1}=0.01$ and $\epsilon_{2}=0.01$. }
	\label{fig:S.10}
\end{figure}

Our study starts considering obstacles with the same amplitude ($\epsilon_{1}=0.01$ and $\epsilon_{2}=0.01$). Figure \ref{fig:S.10} shows the free surface evolution for $f=-0.1$. At early times, the right obstacle emits depression solitary waves which travel downstream and the formation of an undular bore above the obstacle so does the left one. It is different from what happens in the non-resonant subcritical case, in which a steady state is reached (see Figure \ref{fig:S.4}). We point out that from the collision between the depression solitary waves generated by the left obstacle and the bore over the right bump we have  short waves being radiated upstream. Once the depression solitary waves reach the right obstacle a series of wave collision happens and eventually they escape out.

\begin{figure}[!ht]
	\centerline{\includegraphics{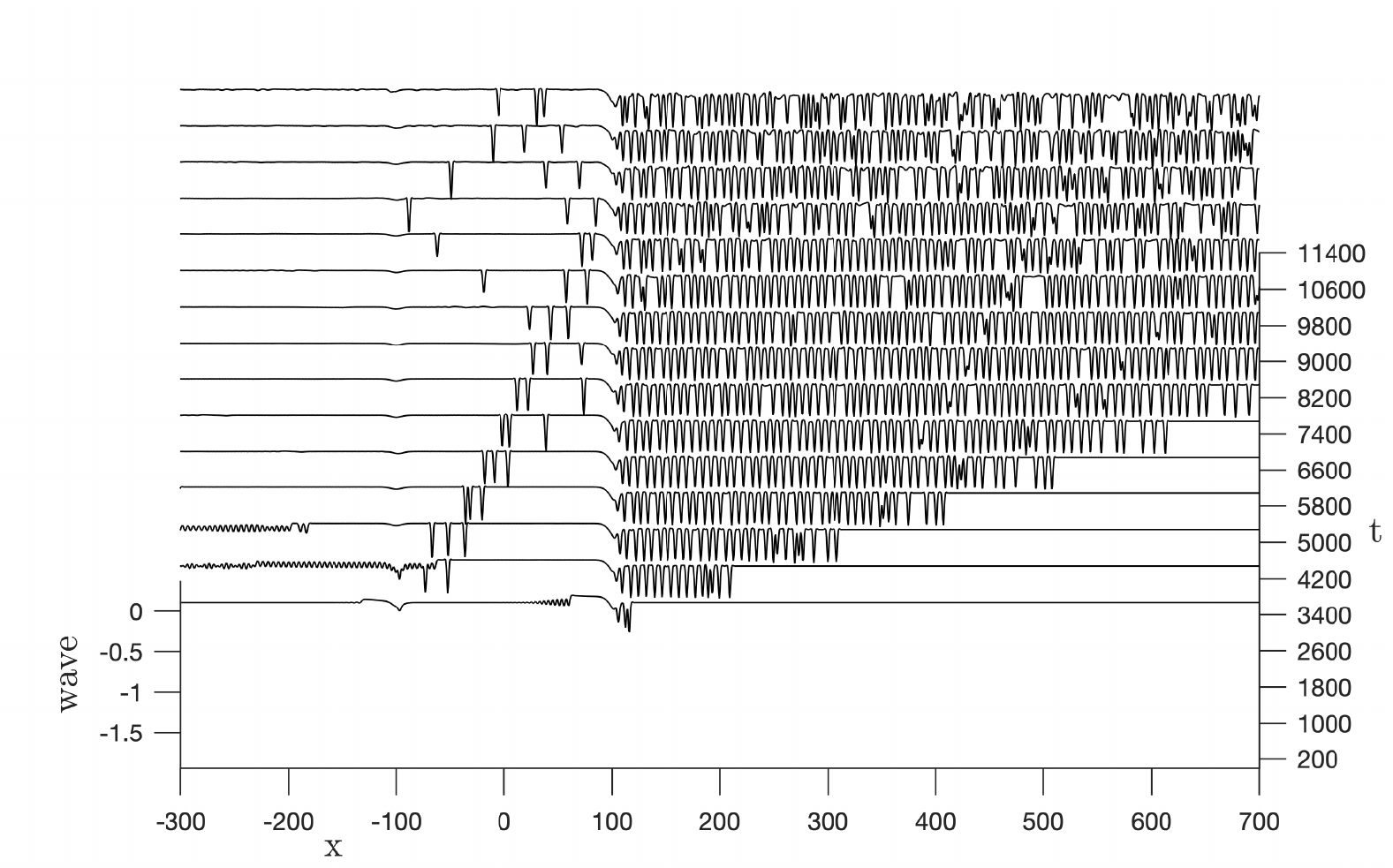}}
	\caption{Subcritical resonant regime free-surface evolution with  $f=-0.1$, $\epsilon_{1}=0.01$ and $\epsilon_{2}=0.03$. }
	\label{fig:S.11}
\end{figure}

\begin{figure}[!ht]
	\centerline{\includegraphics{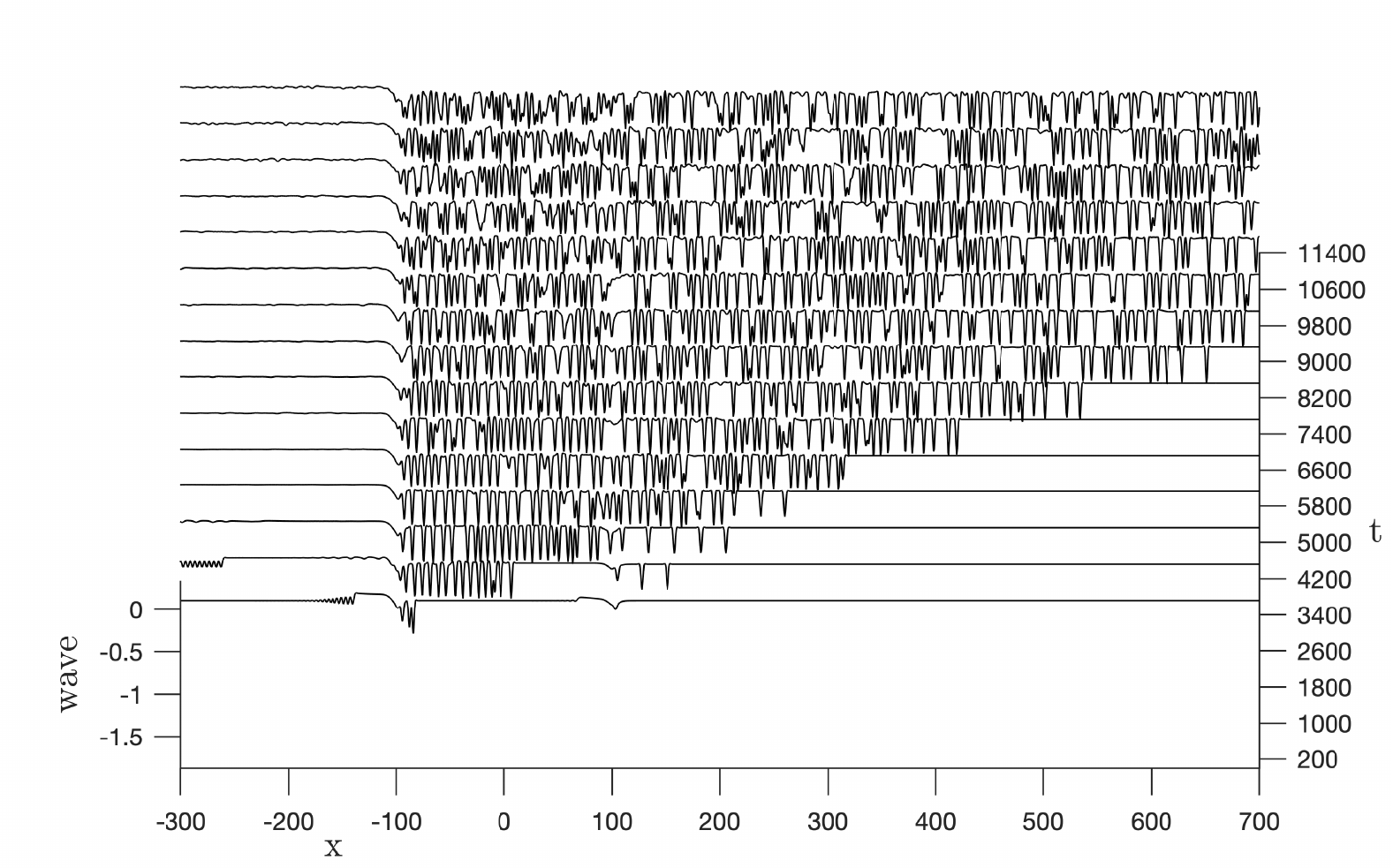}}
	\caption{Subcritical resonant regime free-surface evolution with  $f=-0.1$, $\epsilon_{1}=0.03$ and $\epsilon_{2}=0.01$.  }
	\label{fig:S.12}
\end{figure}

For bumps with amplitudes $\epsilon_{1}=0.01$ and $\epsilon_{2}=0.03$  an interesting scenario arises as can be seen in Figure \ref{fig:S.11}. At early times, the left obstacle emmanates depression solitary waves which travel downstream, then this mechanism ceases. Meanwhile, depression solitary waves are being emitted frenetically  from the second obstacle producing a series of collisions. It is worth noting that as the time elapses the depression solitary waves generated by the left obstacle remain trapped bouncing between the bumps. When $\epsilon_{1}=0.03$ and $\epsilon_{2}=0.01$ both obstacles generate depression solitary waves  periodically with the larger one more rapidly. Since the flux of wave generation of the large obstacle is high, wave collisions highlights the dynamic. More details of this case is given in Figure \ref{fig:S.12}.

\section{Conclusion}
In this paper we have investigated capillary-gravity flows over two obstacles.  Through a pseudospectral numerical method, we showed that the flow  is not necessarily  governed by the larger obstacle as it is in the absence of surface tension. In the supercritical near-resonant case, the flow is mainly described by the formation of an undular bore. In certain regimes, the undular bore generated by both obstacles collide and its structure colapses. In the subcritical near-resonant case, the flow is mainly characterized by the generation of depression solitary waves periodically  that propagate downstream. Besides, we also found cases  in which depression solitary waves remain trapped between the bumps bouncing back and forth. 

\section{Acknowledgements}

The authors are grateful to IMPA-National Institute of Pure and Applied Mathematics for the research support provided during the Summer Program of 2020. M.F. is grateful to Federal University of Paran{\' a} for the visit to the Department of Mathematics. R.R.-Jr is grateful to University of Bath for the extended visit to the Department of Mathematical Sciences.

\bibliographystyle{abbrv}

\begin{thebibliography}{99}
	
      \bibitem[Akylas (1984) ]{Akylas}
	\textsc{Akylas, TR.} 1984
	On the excitation of long nonlinear water waves by a moving pressure distributions.
	{\it J Fluid Mech}. \textbf{141}, 455-466
	
	\bibitem[Baines (1995) ]{Baines}
	\textsc{Baines, P.} 1995
	Topographic effects in stratified flows.
	{\it Cambridge University Press}.
	
	
	\bibitem[Chardard et al. (2011) ]{Chardard}
	\textsc{Chardard, F., Dias, F., Nguyen. HY. \& Vanden-Broeck. JM.} 2011
	 Stability of some stationary solutions to the forced KdV equation with one or two bumps.
	{\it  J Eng Math.},  {\bf 70}, 175-189.

        \bibitem[Ermakov \& Stepanyants (2019) ]{Ermakov}
	\textsc{Ermakov, E. \&   Stepanyants, Y.} 2019
	 Soliton interaction with external forcing within the Korteweg-de Vries equation.
	{\it  Chaos},  {\bf 29}, 013117-1-013117-14.




\bibitem[Falcon et. al (2002)]{Falcon}
	\textsc{Falcon, E., Laroche, C.,  Fauve, S.} 2002
	Observation of depression solitary surface waves on a thin fluid layer
	{\it Phys. Rev. Lett.}, {\bf 89}, 204501-1-204501-4
%% O termo dispersivo desaparece quando B=1/3 fKdV

	\bibitem[Flamarion et al.  (2019)]{Marcelo-Paul-Andre}
	\textsc{Flamarion, M. V., Milewski, P. A. \& Nachbin A.} 2019 
	Rotational waves generated by current-topography interaction.
	{\it Stud Appl Math}, {\bf 142}, 433-464.

 



	
	\bibitem[Grimshaw \& Smyth (1986)]{Grimshaw86}
	\textsc{Grimshaw, R. \& Smyth, N.} 1986
	Resonant flow of a stratified fluid over topography
in water of finite depth.
	{\it J. Fluid Mech.}, {\bf 169}, 429-464.
	
	
      \bibitem[Grimshaw \& Malewoong (2016)]{Grimshaw16}
	\textsc{Grimshaw, R. \& Malewoong, M.} 2016
	Transcritical flow over two obstacles: forced Korteweg-de Vries framework
	{\it J. Fluid Mech.}, {\bf 809}, 918-940.

	
	\bibitem[Grimshaw \& Malewoong (2019)]{Grimshaw19}
	\textsc{Grimshaw, R. \& Malewoong, M.} 2019
	Transcritical flow over obstacles and holes: forced Korteweg-de Vries framework
	{\it J. Fluid Mech.}, {\bf 881}, 660-678.

       \bibitem[Hanazaki et. al (2017)]{Hirata}
	\textsc{Hanazaki, H., Hirata, M., Okino, S. } 2017
	Radiation of short waves from the resonantly excited capillary-gravity waves.
        \textit{J. Fluid Mech.} \textbf{810}, 5-24.

	
	
	
        \bibitem[Joseph (2016))]{Joseph}
	\textsc{Joseph. A.} 2016 
	{Investigating Seaflaws in the Oceans}
	\textit{New York: Elsevier}.
		
	\bibitem[Johnson (2012)]{Johnson}
	\textsc{Johnson, R. S.} 2012 
	{Models for the formation of a critical layer in water wave propagation.}
	\textit{Phil. Trans. R. Soc. A} \textbf{370}, 1638-1660.
	
	
		\bibitem[Lee \& Whang (2015)]{Lee1}
	\textsc{Lee, S. \& Whang, S.} 2015 
	{Trapped supercritical waves for the forced KdV equation with two bumps.}
	\textit{Appl. Math. Model.} \textbf{39}, 2649-2660.
	
	
		    \bibitem[Malomed \&  Vanden-Broeck (1996)]{Malomed}
	\textsc{Malomed, B. \& Vanden-Broeck, J-M. } 1996
	Solitary wave interaction for the fifth-order KdV equation
        \textit{Contem. Mathematics} \textbf{200}, 133-143.
        
	
	
	
	
	
	\bibitem[Milewski (2004)]{Paul}
	\textsc{Milewski, P. A. } 2004
	The Forced Korteweg-de Vries Equation as a Model for Waves Generated by Topography.
	{\it CUBO A mathematical Journal} {\bf 6} (4), 33-51.


       \bibitem[Milewski et. al (1999)]{Milewski3}
	\textsc{Milewski, P. A. \& Vanden-Broeck, J-M. } 1999
	Time dependent gravity-capillary flows past an obstacle
        \textit{Wave Motion} \textbf{29}, 63-79.
	


	

	
	

%	\bibitem[Pratt (1984)]{Pratt}
%	\textsc{Pratt, LJ. } 1984
%	On nonlinear flow with multiple obstructions.
%	{\it J. Atmos. Sci.} {\bf 41} , 1214-1225.
	
	
	
%	
%	\bibitem[Ribeiro-Jr et al. (2017)] {JFM17}
%	\textsc{Ribeiro Jr R., Milewski P.A. \& Nachbin A.}  2017
%	{Flow structure beneath rotational water waves with stagnation points}.
%	\textit{J. Fluid Mech.} \textbf{812}, 792-814.
	
	
	

	
	

              \bibitem[Peyrard (1994))]{Peyrard}
	\textsc{Peyrard, M.} 1994
	{Nonlinear excitations in biomolecules}
	\textit{Berlim: Springer}.
		
		
	\bibitem[Trefethen (2001)]{Trefethen}
	\textsc{Trefethen, L. N.} 2001
	Spectral Methods in MATLAB.
	{\it Philadelphia: SIAM.}
	
	
	\bibitem[Wu (1987)]{Wu1}
	\textsc{Wu, T. Y.} 1987
	Generation of upstream advancing solitons by moving disturbances
	{\it J Fluid Mech}  {\bf 184} , 75-99.
	
	\bibitem[Wu \& Wu (1982)]{Wu2}
	\textsc{Wu. DM \& Wu. T. Y.} 1982
	Three-dimensional nonlinear long waves due to moving surface pressure. In: Proc. 14th. Symp.
on Naval Hydrodynamics
	{\it Nat. Acad. Sci., Washington, DC}, 103-25.
	


\bibitem[Zhu (1995)]{Zhu}
	\textsc{Zhu, Y.} 1995
	 Resonant generation of nonlinear capillary-gravity waves
	{\it Phys. Fluids}, {\bf 7}, 2294-2296.
	
%	
%	\bibitem[Whitham (1974)]{Whitham}
%	\textsc{Whitham, G. B.}  1974
%	Linear and Nonlinear Waves.
%	{\it New York: John Wiley and Sons.}
%	
	
	
	
	
	
	
\end{thebibliography}

\end{document}